\documentclass[12pt,showpacs]{revtex4-1}
\usepackage[version=3]{mhchem}
\usepackage{graphicx}
\usepackage{subfigure}
\usepackage{natbib}
\usepackage[T1]{fontenc}
\usepackage[utf8]{inputenc}
\usepackage[breaklinks=true]{hyperref}
\hypersetup{colorlinks=true, citecolor=blue, urlcolor=blue, linkcolor=blue}

\begin{document}
\title{\textbf{\large{Proliferation of metallic domains caused by inhomogeneous heating near the electrically-driven transition in  VO$_2$ nanobeams}}}
\author{Sujay Singh$^{1}$}
\author{Gregory Horrocks$^{2,3}$}
\author{Peter M Marley$^{2,3}$}
\author{Zhenzhong Shi$^{1}$}
\author{Sarbajit Banerjee$^{2,3}$}
\author{G. Sambandamurthy$^{1}$}
\email{sg82@buffalo.edu}
\affiliation{$^{1}$Department of Physics}
\affiliation{$^{2}$Department of Chemistry, University at Buffalo, State University of New York, Buffalo, NY 14260, USA}
\affiliation{$^{3}$Department of Chemistry, Texas A\&M University, College Station, TX 77843, USA}

\begin{abstract}
We discuss the mechanisms behind the electrically driven insulator-metal transition in single crystalline VO$_2$ nanobeams. Our DC and AC transport measurements and the versatile harmonic analysis method employed show that non-uniform Joule heating causes phase inhomogeneities to develop within the nanobeam and is responsible for driving the transition in VO$_{2}$. A Poole-Frenkel like purely electric field induced transition is found to be absent and the role of percolation near and away from the electrically driven transition in VO$_{2}$ is also identified. The results and the harmonic analysis can be generalized to many strongly correlated materials that exhibit electrically driven transitions.

\end{abstract}

\maketitle
Vanadium dioxide (VO$_{2}$) is a well-studied strongly correlated material that shows a sharp insulator to metal transition (IMT) with a $T_C$ $\sim$ 342 K \cite{MorinPRL59}, accompanied by orders of magnitude changes in both its electrical resistivity and optical transmission. Although the IMT in VO$_{2}$ has been studied for decades, the nature of the transition is still debated and it is believed to exhibit signatures of both Mott correlation and Peierls distortion \cite{WentzcovitchPRL94,RicePRL94,QazilbashScience07}. In addition to being thermally-driven, the IMT in VO$_{2}$ can also be triggered by voltage, light, and strain, which makes it a material of interest for future technological applications such as ultra-fast switches, optical devices, Mott field effect transistors, etc. \cite{YangARMR12,NakanoNature12}. The electrically driven resistive switching is a complicated and dynamical process wherein various factors such as Joule heating, electric field, percolation, oxygen vacancies, and strain can each have influence on the properties \cite{XZhongJAP11,JKimAPL10,AZimmersPRL13,StefanovichJP2000,SleeNatMat08,GopalakrishnanJMatSc09,Tai-LungPRB11,BWuPRB11,ZYangJAP11,JustinNatNano14}. Recent work on realizing electrical switching devices using strongly correlated materials have discussed the importance of Joule heating near the transition \cite{XZhongJAP11,JKimAPL10,AZimmersPRL13} as opposed to a purely electric field induced transition \cite{StefanovichJP2000,SleeNatMat08,GopalakrishnanJMatSc09,Tai-LungPRB11,BWuPRB11,ZYangJAP11}. Recently, the Poole-Frenkel effect (an electric field induced effect) \cite{FrenkelPR38,SimmonsPR67} was observed as a precursor to Joule heating dominated switching in a related material, V$_2$O$_3$ \cite{JustinNatNano14}.  

In the present work, we have studied, by DC and AC transport measurements, the possible underlying mechanisms behind the electrically driven IMT in individual nanobeam devices of single crystalline VO$_2$ (see Supplemental Material: Section 1). Our results are summarized as follows: extremely abrupt IMT was observed at electric fields two orders of magnitude smaller than estimated by a purely Poole-Frenkel type transition. The calculated average temperatures of the nanobeams at the IMT, based on our model, were lower than $T_C$, implicating non-uniform current paths leading to phase inhomogeneity as possible suspects in driving the IMT. To verify the roles of individual mechanisms, a novel harmonic analysis method of the AC transport data was carried out and the results show that Joule heating plays an important role near the transition and Poole-Frenkel effect is strikingly absent. In particular, our measurements provide experimental evidence that phase inhomogeneities exist within the sample and these are verified by both DC and  AC transport measurements. Finally, it appears that the transport behavior of insulating VO$_2$ nanobeams far below the critical voltage is similar to a random resistor network and shows signatures of Joule heating induced percolation. However, deviation from percolation behavior is observed as the threshold voltage for IMT is approached.
\begin{figure}[h]
\centering
\includegraphics[height=10cm]{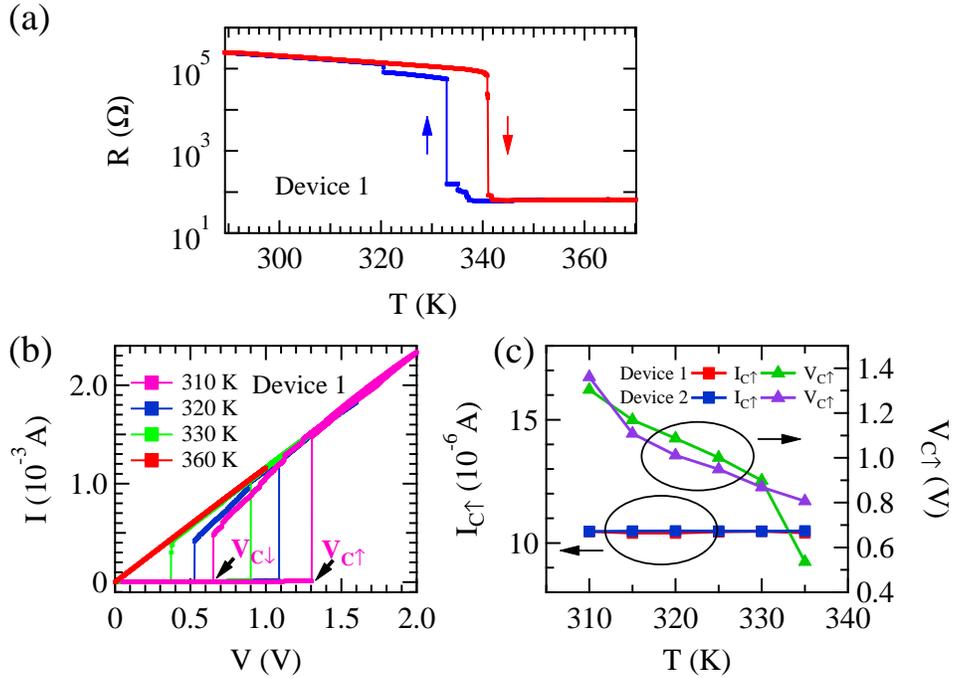}
\caption{(a) Resistance (R) vs. temperature (T) plot showing the IMT at $T_C$ = 342 K while heating. (b) Current-voltage characteristics showing switching at $V_{C\uparrow}$ while sweeping the voltage up. (c) $I_{C\uparrow}$ (left-axis) and $V_{C\uparrow}$ (right axis) as a function of temperature. $I_{C\uparrow}$ is constant at all set temperatures whereas $V_{C\uparrow}$ shows an exponential T-dependence in the measured temperature range.}
\label{Fig1}
\end{figure}

Fig. \ref{Fig1} (a) presents the resistance (R) of a single nanobeam device as a function of temperature (T): the device undergoes a sharp insulator to metal transition around 342 K with orders of magnitude change in the resistance at $T_C$ \cite{GregACS14}. Hysteretic behavior is observed upon cooling to the insulating phase with the switching occurring at 333 K. The insulating side of the R-T trace is fitted to $R=R_{0}e^{\left(\frac{E_{a}}{k_{B}T}\right)}$, where $E_a$ is the activation energy (0.30$\pm$0.03 eV in our devices). 

Having observed the transport behavior in the thermally driven case, next we turn our attention to the electrically driven case. Fig. \ref{Fig1} (b) shows the current (I) as a function of voltage (V) at various set temperatures ($T_S$) measured during the heating part of the R-T cycle. Initially, the current through the device increases smoothly, but non linearly with increasing voltage. At the critical voltage ($V_{C\uparrow}$) the current jumps by orders of magnitude signaling the switch to a highly conducting state. As the bias voltage is lowered from the high conducting state, the device returns to the insulating state wherein a large drop in current at $V_{C\downarrow}$ is preceded by smaller drops. The $V_{C\uparrow}$ values in our devices range from 1.2-2.0 V corresponding to electric fields of 0.24-0.40 V/$\mu$m at 310 K. These values are two orders of magnitude smaller than the critical field values expected from a purely electric field induced IMT in VO$_2$ \cite{KoAPL08,HormozSSE10}. It can be seen in Fig. \ref{Fig1} (b) that the abruptness of the IMT diminishes with increasing $T_S$ \cite{AdamarXiv14} and the jump size and the hysteresis widths were largest at 310 K. Fig. \ref{Fig1} (c) shows the behavior of the critical current ($I_{C\uparrow}$) and critical voltage ($V_{C\uparrow}$) from two devices at various temperatures. The $I_{C\uparrow}$ at the onset of IMT was found to be constant from 335 K to 310 K whereas $V_{C\uparrow}$ decreases with increasing $T_S$ and shows an exponential T-dependence ($V_{C\uparrow} \propto e^{-T/T_0}$) \cite{Tai-LungPRB11}. 

To understand the microscopic mechanisms relevant near the electrically driven switching, we proceed by assuming Joule heating to be the only cause for the non-linearity of the IV characteristics prior the onset of the IMT in Fig. \ref{Fig1} (b). We then assume a current-voltage (IV) relationship, $v=i\left(R_B+R_{0}e^{\left(\frac{E_a}{k_B(T_{S}+\Delta{T})}\right)}\right)$ based on the thermally activated behavior of resistance and taking the current density and temperature distribution to be uniform across the nanobeam ($R_{B}$ is the resistance in series to the nanobeam device). In this picture, the non-linear behavior in the IV characteristics is only due to a change in the temperature and hence Ohm's law is not violated. The IV relation includes the contribution $\Delta{T}$ due to Joule heating and can be further simplified using Taylor series approximation to $i\approx\frac{v}{R_{B}+R_{S}e^{\left(\frac{-\Delta{T}}{T_{0}}\right)}}$ (see Supplemental Material: Section 2). It is clear from above that the IV characteristics would be linear when Joule heating is insignificant and hence $\Delta{T}$ will be negligibly small. On the other hand, introduction of Joule heating introduces non-linear behavior in the IV characteristics. An average temperature raise ($\Delta{T}$) from the Joule heating effect using this picture can be estimated as $\Delta{T}=-T_{0}\ln\left[\frac{1}{R_{S}}\left(\frac{v}{i}-R_{B}\right)\right]$, where $R_{S}$ is the resistance at $T_S$ and $T_{0} = k_{B}T_{S}^2/E_{a}$. 
\begin{figure}[t=h]
\centering
\includegraphics[height=10cm]{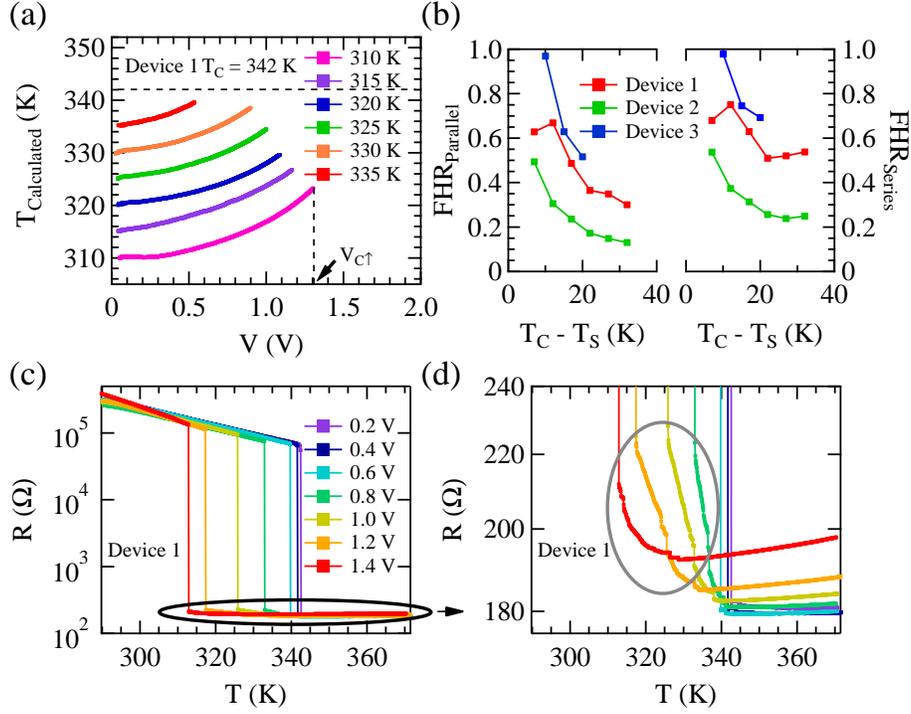}
\caption{(a) Average calculated temperature at several $T_{S}$ values (below $T_{C}$) leading up to $V_{C\uparrow}$. (b) Fraction of hot region (FHR) based on parallel (left axis) and series (right axis) hot and ambient regions as a function of $(T_C-T_S)$. (c) R-T traces measured at various voltage bias values for device 1. (d) Enlarged portion of Fig. \ref{Fig2} (c), gray oval highlights the insulating-like behavior even after the switching.}
\label{Fig2}
\end{figure}

Fig. \ref{Fig2} (a) shows the calculated average temperature in the nanobeam until the onset of IMT. When the set temperature of the nanobeam were 330 K and 335 K, the average temperature of the nanobeam were close to $T_{C}$ at $V_{C\uparrow}$. However, the average temperature of the nanobeam was far from $T_C$ at $V_{C\uparrow}$, when IV characteristics were measured at lower temperatures. The temperature profile inside the nanobeam due to Joule heating likely depends on experimental conditions such as $T_{S}$, electric field, device geometry, the thermal coupling to the substrate, electrodes, and the dimensions of the nanobeam \cite{LuRSI01}. The temperature estimation in Fig. \ref{Fig2} (a) is not based on power dissipation in the nanobeam, however including factors such as power dissipation and thermal coupling will only lower the estimated temperature raise. Since the estimation of temperature raise in the nanobeam is temperature based on the assumption that the device has uniform current density and homogeneous temperature distribution, the deviation of the average calculated temperature from $T_C$ at the IMT is seemingly due to the non-uniform conduction that can arise from coexistence of metallic and insulating domains. 

Furthermore, a part or all of the nanobeam need not reach $T_C$ in order for a macroscopic observation of the IMT in transport measurements. Other contributing factors such as local disorder and/or microscopic strain distribution may serve as nucleation sites for domains and switching can occur. However, a strong case can be made for non-uniform temperature distribution within the nanobeam since the average temperature increase ($\Delta{T}$) due to thermal heating was less than the temperature needed ($T_C-T_S$) to reach $T_C$.  In VO$_2$, metallic and insulating domains are known to exist in parallel as well as in series combinations along the length of the nanobeam \cite{GuNL07,OkimuraJJAP09,EFAPL13,BSMunAPL13,UedaAPL13}. Using a simplistic model that assumes parallel and series combination of hot and ambient region, we have estimated the fraction of Joule heating induced hot region (FHR) in the nanobeam at the onset of IMT. FHR$_{parallel}$ is given by a ratio of the areas of the cross section of the hot region and the total cross-sectional area whereas FHR$_{series}$ is the ratio of the lengths of the hot region to the total length (see Supplemental Material: Section 3). 

The left and right axes of Fig. \ref{Fig2} (b) show FHR for the parallel and series cases respectively as a function of the temperature raise needed ($T_C-T_S$) to drive IMT. It shows that the extent of hot regions induced due to Joule heating in the nanobeam decreases with the decrease in $T_S$ for both parallel and series case. If a conducting channel indeed is formed, the nanobeam will not be completely in metallic phase above the switching and some remnant insulating regions are likely to be present. To verify this scenario, we have measured R-T with different voltage bias values starting from 50 mV to 1.4 V (Fig. \ref{Fig2} (c)). $T_C$ is found to decrease with an increase in voltage bias due to the Joule heating effect. It is interesting to note from  Fig. \ref{Fig2} (d) that the resistance continued to decrease with temperature even after the large drop, signaling remnant insulating behavior beyond switching and is suggestive of a non-uniform electronic phase within the nanobeam.

In order to further identify the role of Joule heating on the onset of electrically driven IMT, we focused on understanding the non-linearity of the IV characteristics up to $V_{C\uparrow}$ by employing a harmonic analysis of the AC signal across the device. Typically, any non-linear IV relation can be expressed as a power series $I = \sum\limits_{i=0}^nk_{i}V^i$, where coefficient $k_{i}$ is related to the power of an individual harmonic \cite{Solbach82,Cullen82}. The magnitude of $k_{i}$ can be estimated by measuring harmonics present in the AC voltage. The method of harmonic detection and their comparison have been employed previously to understand the effect of second harmonic in microwave characteristics of Schottky barrier diodes, to understand percolation and breakdown in semicontinuous metal films, and to measure thermal conductivity of various systems \cite{Solbach82,Cullen82,WrightPRB86,DubsonPRB89,CahillRSI90,YagilPRB92,YagilPRL92,LuRSI01,DamesRSI05}. In our case, if the resistance of the device is a Poole-Frenkel like function of the electric field which is analogous to the Schottky effect \cite{FrenkelPR38,SimmonsPR67}, then the second harmonic along with all other harmonics should be present in the AC electrical signal \cite{Solbach82,Cullen82}. Based on Taylor series approximation, the magnitude of electric field generated second harmonic is expected to be the strongest after the fundamental frequency. On the other hand, if Joule heating plays a dominant role, then power and temperature will oscillate with $2\textit{f}$ ($\textit{f}$=bias frequency) that will lead to the resistance oscillation giving rise to the presence of third harmonic in the AC signal across the device \cite{WrightPRB86,DubsonPRB89,CahillRSI90,YagilPRB92,YagilPRL92,LuRSI01}. The third harmonic generated by Joule heating will be the strongest after the fundamental frequency and hence the harmonic analysis of the signal across the device can potentially provide clues about the relevant mechanisms. 
\begin{figure}[th]
\centering
\includegraphics[height=10cm]{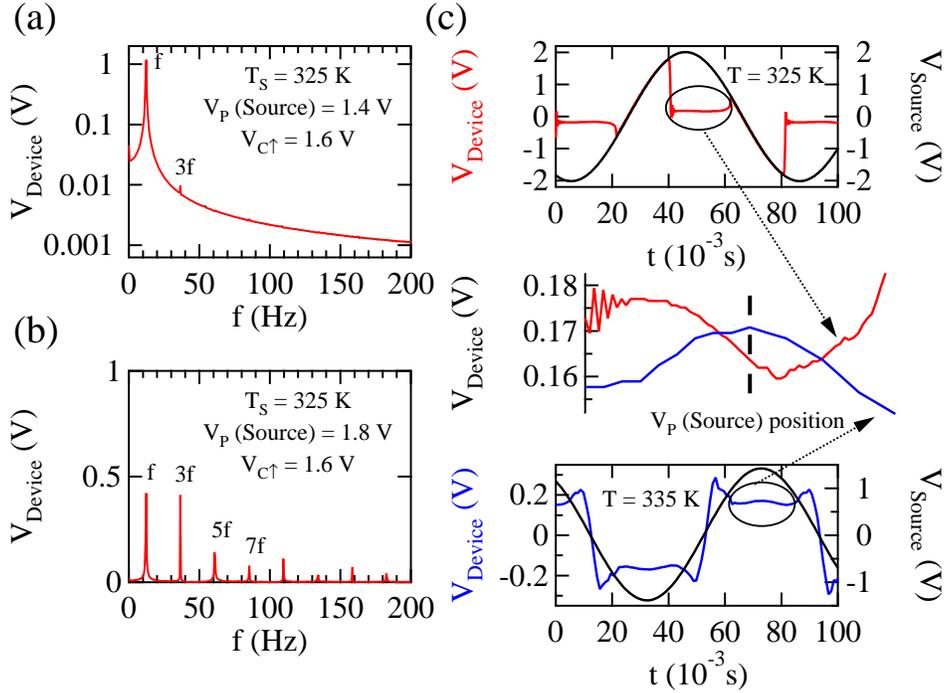}
\caption{(a), (b) The response of a device in the frequency domain at $V_{P} (Source)<V_{C\uparrow}$ and $V_{P} (Source)>V_{C\uparrow}$ respectively using a source frequency of 12.3433 Hz. (c) AC signal across the device in time domain at 325 K and 335 K; middle panel shows the device responses in the high conducting phase.}
\label{Fig3}
\end{figure}

Fig. \ref{Fig3} (a) shows the response in the frequency domain recorded from an insulating nanobeam (T = 325 K) when an AC source voltage was applied.  A peak at the third harmonic of the applied bias frequency was observed and its magnitude increased with increasing bias voltage; however no discernible peak at the second harmonic was observed. This clearly shows that effect of electric field expected from a Poole-Frenkel like mechanism is insignificant on the observed transport behavior near the IMT. Fig. \ref{Fig3} (b) shows the frequency response of a device when the peak value of the source is higher than $V_{C\uparrow}$. In this case, the device oscillates between low and high conducting states leading to the generation of several odd harmonics. It is interesting to note that VO$_{2}$ has the potential to be used as a material for harmonic generation as even the magnitude of the $19^{th}$ harmonic after switching was found to be $1/10^{th}$ of the magnitude of the fundamental frequency. 

The presence of a conducting channel as postulated in the previous section can also be verified by comparing the time domain signals of the source with the signals across the device at different $T_S$. Fig.  \ref{Fig3} (c) shows that at 325 K, in the high conducting state, the device voltage shows transient oscillations after switching and does not track the source indicating phase inhomogeneity and possible coexistence of insulating and metallic regions.  However, the device voltage does indeed follow the source in the high conducting state at 335 K pointing to the presence of a single phase within the nanobeam.
 
Once the presence of phase inhomogeneity is established, the natural next step is to look for signatures of Joule heating induced percolation in the transport measurements. It has been previously shown that the third harmonic of AC signal from the device can be thought of as the fourth moment of the current distribution and can provide information about percolation due to local Joule heating similar to information obtained from 1/f noise measurements \cite{WrightPRB86,DubsonPRB89,YagilPRB92,YagilPRL92}. According to this model \cite{WrightPRB86,DubsonPRB89,YagilPRB92,YagilPRL92}, the third harmonic coefficient ($B_{3\textit{f}}$) scales as, $B_{3\textit{f}}\propto R^{2+\textit{w}}$ where $B_{3\textit{f}}=V_{3\textit{f}}/I_{0}^3$ and $\textit{w}$ is the critical exponent. $B_{3\textit{f}}$ will be linearly proportional to $R^{2}$ in the case when there is no change in the current distribution in the system. For a random resistor network, $B_{3\textit{f}}$ scales with resistance (R) and critical exponent ($\textit{w}$) values between 0.8 and 1.05 have been observed in earlier experiments \cite{WrightPRB86}. 

\begin{figure}[h]
\centering
\includegraphics[height=8cm]{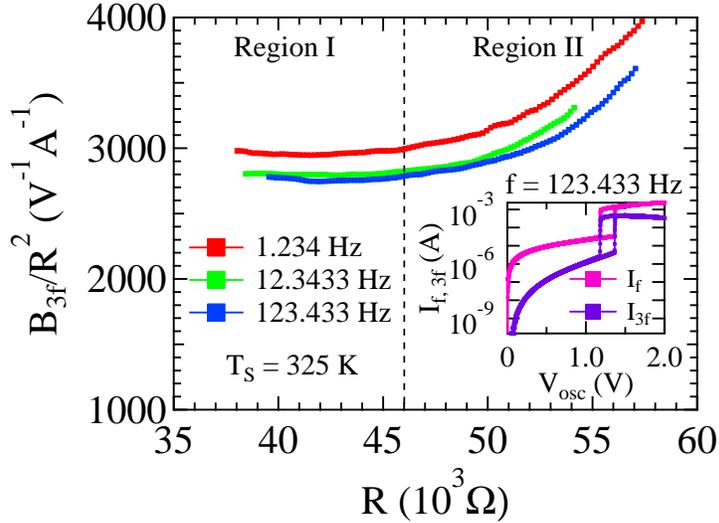}
\caption{B$_{3\textit{f}}$/R$^{2}$ vs. R at various source frequencies. The nanobeam behaves as a random resistor network ($\textit{w}=1.2-1.5$) at low bias (region II) and deviation from the behavior is evident in the high bias region (region I) where exponent $\textit{w}=0$. The inset shows the behavior of first and third harmonics of the device current.}
\label{Fig4}
\end{figure} 

If our VO$_{2}$ device has a random distribution of insulating and metallic domains, then $\textit{w}$ should have a value similar to that of a random resistor network. If there is no percolation due to Joule heating or no change in the connectivity between domains then $\textit{w}$ will be zero. When the IVs were measured using lock-in amplifiers both the first and the third harmonics were simultaneously recorded while sweeping the source voltage (inset of Fig. \ref{Fig4} shows the IV characteristics of the first and third harmonics). Fig. \ref{Fig4} shows the evolution of a scaled third harmonic coefficient ($B_{3\textit{f}}/R^{2}$) as a function of device resistance in the insulating phase. We have roughly marked the behavior into two regions: in region I (low resistance region, closer to the transition), the nanobeam does not behave as a classical random resistor network as can be seen from the negligible slope of the traces indicating no changes in connectivity or relative current distribution across the nanobeam. However, in the second region (high resistance region, away from the transition) the behavior of ($B_{3\textit{f}}/R^{2}$) resembles that of a random resistor network (with \textit{w}=1.2 to 1.5) which may be thought of as a signature of local percolation between metallic domains. At bias values far below $V_{C\uparrow}$, a random network of resistors is discernible based on the value of the critical exponent deduced from the measurements shown in Fig. \ref{Fig4}; however, subsequently, the percolation pathways are replaced by plausible conducting channels that are likely defined by avalanche-type cascading processes \cite{XZhongJAP11}. Upon these interconnections having been defined, random percolative behavior is no longer observed as can be seen in region I in Fig. \ref{Fig4}. These analyses show that local Joule heating induced percolation does indeed play a role at bias values far below  $V_{C\uparrow}$ and acts as a precursor to the avalanche-type Joule heating processes leading to an extremely abrupt IMT in VO$_2$.

The electric field at the onset of IMT in our devices are two orders of magnitude smaller than the critical electric field estimates based on purely electric field induced switching in VO$_2$. The calculated average temperatures of the nanobeams based on our model are lower than T$_C$, implicating non-uniform current paths leading to phase inhomogeneity as possible suspects. Furthermore, harmonic analysis of the AC electrical signals from the device shows that Joule heating plays a significant role as compared to the electric field in underpinning the electrically driven IMT in VO$_2$. It appears that the transport behavior of insulating VO$_2$ nanobeams is similar to a random resistor network behavior at bias values far below the $V_{C\uparrow}$. The occurrence of avalanche-type events closer to $V_{C\uparrow}$ likely induces the formation of conducting channels that alter conduction through the nanobeams, thereby precluding further need for percolation. Our measurements provide evidence that phase inhomogeneities exist below the IMT in VO$_2$ and further show the importance of understanding the microscopic mechanisms relevant near phase transitions in correlated oxide nanostructures. The harmonic analysis method employed to understand the role of Joule heating is a versatile technique and can potentially be applied to understand the nature of electrically driven phase transitions in oxide nanostructures that are technologically important. 

This work was supported by the National Science Foundation under DMR 0847324; G. A. H, P. M, and S.B. acknowledge support from the National Science Foundation under IIP 1311837. 

\end{document}